\documentclass{emulateapj}

\slugcomment{The Astrophysical Journal, Vol. 637: L57-L60, 2006 January 20}
\shorttitle{Optically resolved debris disks at 1 Gyr}
\shortauthors{Kalas et al.}

\begin{document}

\title{First scattered light images of debris disks around HD 53143 and HD 139664}

\author{Paul Kalas\altaffilmark{1}, James R. Graham\altaffilmark{1}, Mark C. Clampin\altaffilmark{2}, \& Michael P.
Fitzgerald\altaffilmark{1}}
\affil{}

\altaffiltext{1}{Astronomy Department, University of California,
Berkeley, CA 94720}
\altaffiltext{2}{Goddard Space Flight Center, Greenbelt, MD  20771}

\begin{abstract}
We present the first scattered light images of debris disks around a K star (HD 53143)
and an F star (HD 139664) using the coronagraphic mode of the Advanced Camera for Surveys (ACS)
aboard the {\it Hubble Space Telescope} (HST).  With ages
0.3 - 1 Gyr, these are among the oldest optically detected debris disks. 
HD 53143, viewed $\sim$45$\degr$ from edge-on,
does not show radial variation in disk structure and has width
$>$55 AU.  HD 139664 is seen close to edge-on and
has  belt-like morphology with a dust peak 83 AU from the star and
a distinct outer boundary at 109 AU.
We discuss evidence for significant diversity in the radial architecture of
debris disks that appears unconnected to 
stellar spectral type or age.  
HD 139664 and possibly the solar system belong
in a category of narrow belts 20$-$30 AU wide.  HD 53143 represents a class of
wide disk architecture with characteristic width $>$50 AU.
\end{abstract}

\section{Introduction}
The configuration of our solar system is perhaps the most significant starting
point for our understanding of planet formation.  Therefore a
fundamental question is whether or not the architecture of our solar system is
common relative to other planetary systems.  
One point of comparison is the structure of our Kuiper Belt relative to other
systems, which are typically seen as debris disks in scattered light or thermal emission.
In scattered light, some debris disks, such as $\beta$ Pic and 
AU Mic, have central holes, but are radially extended to hundreds of
AU radii \citep{smith84, kalas04}.  Other debris disks, such as HR 4796A and Fomalhaut 
consist of relatively narrow rings with sharp inner and outer
boundaries \citep{schneider99, kalas05a}.  
However, a narrow-belt architecture has not previously been
detected in scattered light among stars similar in spectral type and age to the Sun.

HD 53143 (K1V) and HD 139664 (F5V) are two stars $\sim$18 parsec from the Sun 
known to have circumstellar dust due to excess thermal emission at far-infrared 
wavelengths (Aumann 1985; Stencel \& Backman 1991; Table 1).  
Various indicators place the age of 
HD 53143 at $1.0 \pm 0.2$ Gyr \citep{decin00, song00, nord04}, whereas
HD 139664 may be a younger system with age  $0.3^{+0.7}_{-0.2}$ Gyr \citep{lach99, montes01, mallik03, nord04}.
For these two stars the infrared excess corresponds
to a dust mass 3 - 10 times smaller than that of $\sim$10 Myr-old systems such
as AU Mic and $\beta$ Pic \citep{kalas04}.  
Direct imaging of debris disks with masses this small is observationally
challenging, but it is now feasible using the optical coronagraph in ACS.

\section{Observations \& Data Analysis}
We utilized the HST ACS High Resolution Camera (HRC)
with a 1.8$\arcsec$ diameter 
occulting spot to artificially eclipse each star (Table 1; Fig. 1).  
Five F stars were observed in consecutive orbits, as were five K stars, 
in order to minimize differences in the point spread function (PSF) due to telescope thermal variations.
Each PSF was subtracted using the four other stars in 
each set of observations.  
The relative intensity scaling and registration between images was iteratively 
adjusted until the residual image showed a mean radial profile equal to zero 
intensity.  

After the excess nebulosity was detected around HD 53143 and HD 139664, 
we determined that no surface brightness asymmetries were detected between 
each side of each disk.  To improve the signal-to-noise, we mirror averaged the data (Figs. 2 \& 3).  
Mirror averaging  splits the image into two halves along the axis that is 
perpendicular to the disk midplane and bisects the disk.  One side is transposed 
onto the other side and the data are then averaged.  
Asymmetries due to scattering phase function effects will be coadded \citep{kalas96}.  
In effect mirror averaging doubles the integration time 
spent on the circumstellar disk given that the broad features between each 
side are symmetric.  As a test, we also subtracted the two disk halves from 
each other and confirmed that the assumption of symmetry is valid.  


\section{Results}
The two disks have different morphologies due to different inclinations
and intrinsically different architectures.  To quantify the 
viewing geometries and structural properties of the disks, we produce a series
of simulated scattered-light disks that explore the parameters of inclination to the line of sight, 
inner and outer disk radius, and the radial and vertical
variation of dust number density \citep[Fig. 2;][]{kalas96}.  We reinsert the simulated
disks across each star in a direction orthogonal
to the observed midplanes and select those models that
most closely resemble the properties of the observed disks.  
Table 1 summarizes our findings.

The shape of the midplane surface brightness distribution differs significantly for
each system (Fig. 3).  The HD 53143 midplane surface brightness decreases monotonically with projected radius, approximately
as $r^{-3}$, where $r$ is the projected radius.  In the simulated disk, the
radial number density distribution decreases as $q^{-1}$, where $q$ is radius in the disk cylindrical
coordinate system.  
Our solar system's Zodiacal dust complex has
a comparable dependence of grain number density as a function of
radius ($q^{-1.34}$;  Kelsall et al. 1998), controlled mainly by the force of
Poynting-Robertson (PR) drag that causes small grains to spiral into the Sun \citep{burns79}.
Therefore the HD 53143 disk may simply represent
a population of unseen parent bodies that collisionally replenish dust
that is redistributed radially
by PR  drag.   
Our simulations show that the outer radius of the observed disk is a sensitivity-limited
value at approximately 6$\arcsec$ radius (110 AU).  
The inner radius is also sensitivity-limited to 3$\arcsec$ radius (55 AU).  Therefore, the debris disk
around HD 53143 is at least 55 AU wide.

Material surrounding HD 139664, on the other hand, 
is confined to a narrow belt, as indicated by a turnover
in the midplane surface brightness profile 
between 4.5$\arcsec$ and 5.5$\arcsec$ (79 - 96 AU; Fig. 3).  
Our disk simulations show that the peak in the dust distribution occurs
at 83 AU, decreasing as $q^{-2.5}$ from 83 - 109 AU, and with a sharp
outer truncation at 109 AU (Figs. 2 \& 3).  
We tested model disks that have outer radii $>$109 AU and found that these disks would have been 
detectable in our data as far as 10$\arcsec$ radius (175 AU; Fig. 3).  
Given an absence
of significant gas, the belt-like nature of HD 139664 is most likely 
a structure related to planet formation.  \citet{kenyon04} find that
a dust belt with peak surface brightness at $\sim$80 AU radius forms within a planetesimal disk
at age 400 Myr.  The appearance of this belt signals the recent formation
of a $\geq$1000 km planet that gravitationally stirs planetesimals in
its viscinity.  \citet{liou99} and \citet{moro02}, on the other hand, simulate the concentrations of
dust in trans-Neptunian space that arise due to trapping in mean motion resonances.
A natural explanation for the belt-like morphology of HD 139664
is that the $\sim$83 AU peak in the dust distribution
corresponds to either an interior or exterior mean motion resonance
created by a companion to HD 139664.  Large grains may
dominate the belt's peak at 83 AU, with smaller grains 
passing quickly through the resonance regions due to radiation forces.


\section{Discussion}
Taking a census of eight debris disks resolved in scattered light and the predicted
dust distribution in our Kuiper Belt, 
we observe two basic architectures
that are not correlated with stellar mass and luminosity, 
but must depend on other environmental factors (Table 2).
Debris systems are either narrow belts or wide disks.  Both types have central dust depletions, 
but the distinguishing characteristic
is the presence or absence of a distinct outer edge.  HR 4796A, 
Fomalhaut, HD 139664 and the Sun 
are examples of narrow-belt systems.  The belt systems appear
to have radial widths ranging between 20 and 30 AU, and the inner edges may begin as close as
25 AU (Sun), or as far as 133 AU (Fomalhaut).  
HD 32297, $\beta$ Pic, HD 107146,
HD 53143 and AU Mic  are examples of disks with sensitivity-limited outer edges
that imply disk widths $>$50 AU.  

The F, G, and K stars are interesting because {\it a priori} we might expect to find planetary
systems similar to our own, yet we discover significant diversity in the outer regions that correspond
to Neptune and our Kuiper Belt.
With age $\sim$1 Gyr, HD 53143 is among the oldest known extrasolar debris disks,
yet its wide-disk architecture resembles that of the $\sim$10 Myr old systems
of $\beta$ Pic and AU Mic.   The lingering dust mass (Table 1) throughout the system
could signal the absence of giant planets that otherwise sweep clear the 
parent bodies (comets and asteroids) responsible for the
producing the dust disk.  Yet the presence of a dust disk out to at least 110 AU
radius shows that the primordial circumstellar disk probably contained the prerequisite mass
of gas and dust to form giant planets.  
By contrast, the younger system HD 139664 has already developed
a narrow-belt architecture.  

Narrow-belt architectures for the underlying population of planetesimals 
may originate from early stochastic dynamical events,
such as a close stellar flyby, that strip disk mass and dynamically heat the
surviving disk \citep{ida00, adams01}.  Theoretical simulations show that a reduction in disk mass, 
combined with dynamical heating, produces a less stable planetary system that is more likely to
eject giant planets from their formation site to much larger radii, as has been proposed for
the origin of Neptune \citep{thommes99,tsiganis05}.  Therefore, planetesimal belts not only evolve
into a narrow structures because of external stochastic events, but they may be found
at large distances from the central star due to the subsequent
outward migration of interior planets.  

The collisionally replenished dust population will spread away
from any narrow belt of planetesimals.  If the scattered light appearance continues to manifest
a narrow structure, then both the inner and outer edges are probably maintained by other
gravitational perturbers such as stellar or sub-stellar companions.   However, only HR 4796
has a known stellar companion that may truncate the outer radius of the dust belt \citep{aug99}.
 If there is no confinement
mechanism for the outer radius, then the architecture will manifest as a wide-disk.  For example, 
$\beta$ Pic and AU Mic may have narrow belts of planetesimals \citep{aug01, strubbe05}, but the 
observed dust disk widths extend to hundreds of AU.  
Though the predicted structure of our Kuiper Belt places the solar system
among the narrow disk architectures, it is conceivable that the dust component
extends to greater radii \citep{trujillo01}.  Thus the Sun's classification as a narrow-disk
system is tentative.
%


\section{Summary}

We present the first optical scattered light images of debris disks surrounding
relatively old main sequence F and K stars.
Material around HD 139664 is concentrated at 83 AU radius, with a distinct
outer edge at 109 AU, and a depleted, but not empty, region at $<$83 AU radius.  
Dust surrounding HD 53143 has a monotonic $q^{-1}$ variation in grain number density, and
the disk edges from 55 AU to 110 AU are sensitivity-limited values.  
The different radial widths appear consistent with a more general grouping of
debris disks into either narrow or wide architectures.  These two categories
are probably an oversimplification of significant diversity in the formation and
evolution of debris disks.
Future observations should test for common traits
among these stars, such as stellar multiplicity and the existence of planets.


\acknowledgements
{\bf Acknowledgements:}  Based on observations with the NASA/ESA Hubble Space Telescope obtained
at the Space Telescope Science Institute (STScI), which is operated by the Association of Universities for
Research in Astronomy.  Support for Proposal number GO-9475 was provided by NASA through a grant from STScI
under NASA contract NAS5-26555.







%
%
%


\clearpage


\begin{figure}
\plotone{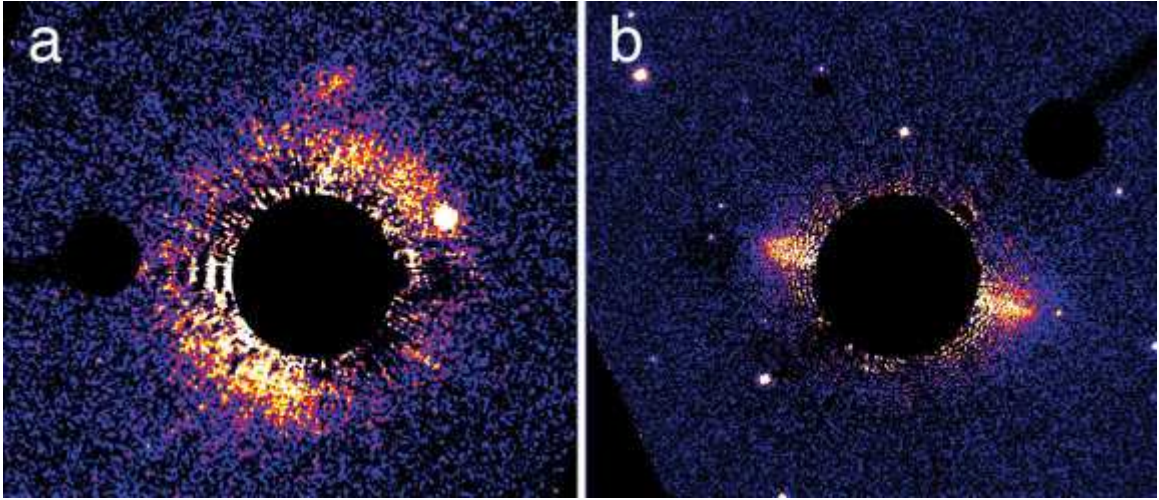}
\caption{
False-color, ACS/HRC F606W ($\lambda_c$= 591 nm, $\Delta\lambda$ = 234 nm)
images of scattered light from debris disks surrounding 
(a) HD 53143 and
(b) HD 139664.  North is up, east is left and the circular black mask in each image 
has radius 3$\arcsec$.  The periphery of each field also shows the 3.0$\arcsec$ diameter 
circular occulting mask located 
at the tip of an occulting bar.  Cumulative integration times are 2340 s and 2184 s for
HD 53143 and HD 139664, respectively.   HD 53143 and HD 139664 were observed on
2004 September 11 and October 14, respectively.
The stellar full-width at half-maximum is 63 mas. 
The images are corrected for distortion and have 25 mas per pixel.   
 \label{fig1}}
\end{figure}

\begin{figure}
\epsscale{0.9}
\plotone{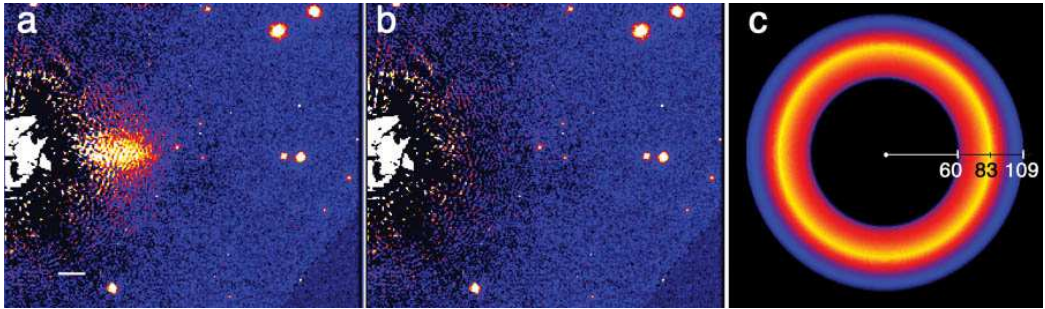}
\caption{
(a)  Mirror averaged image of HD 139664.  
The white bar has length 1$\arcsec$ and the right tip is located at 3$\arcsec$ radius 
from the star (the radius where we begin making measurements 
of disk properties).  (b)  The same data after a model disk has been 
subtracted.  Measurements of the PSF residual along the disk midplane 
are now indistinguisheable from all other position angles and we consider 
the model a satisfactory fit to the data.  
Model disks with inclination range $i = 85\degr - 90\degr$ fit the 
observed morphology and midplane surface brightness.
This range of inclinations is allowed because a vertically fat, edge-on ($i = 90\degr$) disk can have isophote
morphology that resembles a vertically thin disk  with $i = 85\degr$.
(c)  Illustration of the narrow-belt 
architecture by inclining the model HD 139664 disk to face-on. 
Model properties are discussed in Fig. 3.
The dust-free hole within 60 AU must 
be confirmed by future observations.  Summing the light between 82 
and 84 AU radius in our face-on model disk, we obtain $m_v$ = 20.3 mag.  
The perpendicular optical depth is 
approximately $10^{((4.6 - 20.3) / -2.5)} = 5.2\times10^{-7}$ for 
albedo = 1.
 \label{fig2}}
\end{figure}

\begin{figure}
\epsscale{0.75}
\plotone{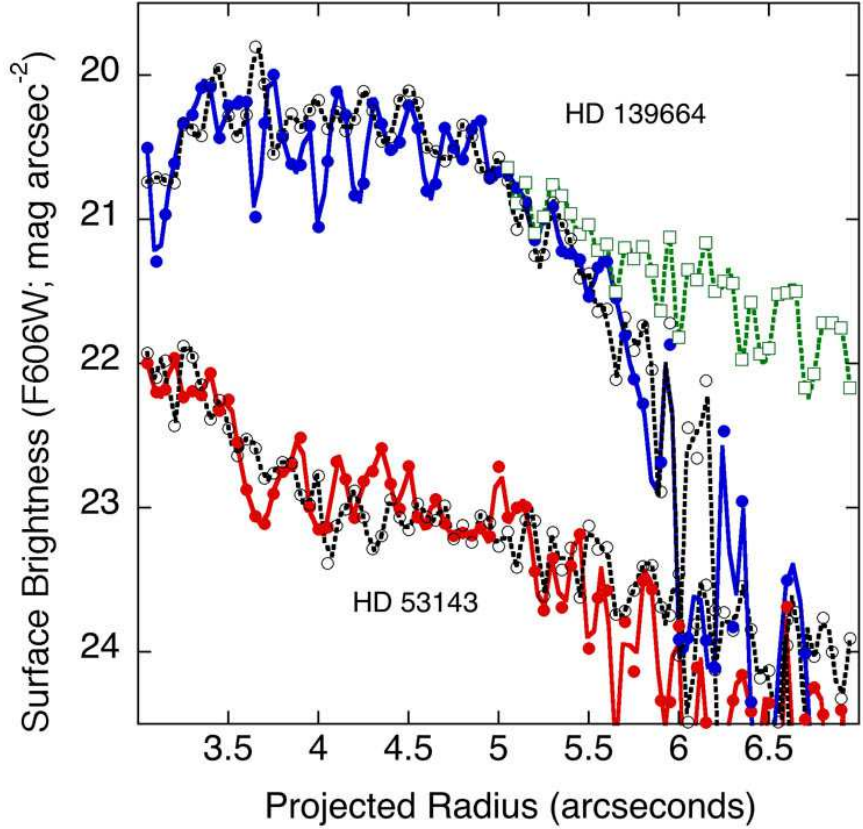}
\caption{
Radial surface brightness distribution along the HD 53143 and HD 139664 disk midplanes. 
Filled circles and solid lines trace the observed disk midplanes, and open circles and dashed lines
trace model dust disks inserted into the data 
orthogonal to the real midplane direction.  The real and model disk midplanes were sampled in
a strip 1.25$\arcsec$ wide for HD 53143, and 0.25$\arcsec$ wide for HD 139664.  The noise
is indicated by the scatter of data points.
A turnover in surface brightness between 4.5$\arcsec$ and 5.0$\arcsec$ (79 - 88 AU) is common to all 
PSF subtractions of HD 139664 (solid blue circles).  
If the HD 139664 disk extended radially inward, then 
the surface brightness profile should be continuously brighter inward.  
In order to match the midplane profile, the model disk (black open circles)
has no dust within 3.40$\arcsec$ radius (59.5 AU).  Between 3.40$\arcsec$ and 4.75$\arcsec$ (59.5 - 83.1 AU) 
the dust density $increases$ as $q^{+3.0}$.  
Between 4.75$\arcsec$ and 6.25$\arcsec$ (83.1 - 109.4 AU), 
the dust number density $decreases$ as $q^{-2.5}$.  The model disk has no dust
beyond 109.4 AU.  
For comparison, a model disk (green open squares) with
outer radius twice that of the best fit case would
be detected 
as far as 10$\arcsec$ radius.  The models for HD 54143, on the other hand, do not 
indicate significant variations in radial structure between the detection limits
of 3$\arcsec$-6$\arcsec$.  All models assume isotropic scattering with scaling
parameters similar to those used for $\beta$ Pic by \citet{kalas95}.
\label{fig3} }
\end{figure}



\clearpage 





%
\clearpage

\begin{deluxetable}{lllll}
\tabletypesize{\scriptsize}
\tablecaption{Star (rows 1-7) and disk (rows 8-15) properties 
\label{tbl-1}}
\tablewidth{0pt}
\tablehead{
\colhead{} & \colhead{HD 53143} & \colhead{HD139664}
}
\startdata
Age (Gyr)							&1.0$\pm$0.2	&$0.3^{+0.7}_{-0.2}$\\
Spectral Type 						&K1V 		& F5IV-V \\
Mass (M$_\odot$)					&0.8			&1.3\\
T$_{eff}$ (K) 						&5224 		& 6653\\
Luminosity (L$_\odot$)				&0.7			&3.3\\
Distance (pc)  						&18.4		& 17.5 \\
$m_V$ (mag)						&6.30		& 4.64 \\
&&\\
Peak disk surf. bright. (mag/arcsec$^{2}$)		&22.0$\pm$0.3	&20.5$\pm$0.3\\
Disk position angle (degrees)					&147$\pm$2	&77 $\pm$ 0.5\\
Inclination (degrees)							&40 - 50		&85 - 90\\
Disk number density gradients	($q^{\alpha}$)			&-1			&+3.0 \& -2.5 \\
Inner dust depletion (AU)						&	$<55$		&83 \\
Maximum outer radius (AU)					&		$>110$	&109  \\
Optical depth from IRAS data\tablenotemark{a}	&$2.5\times10^{-4}$ & $0.9\times10^{-4}$ \\
Optical depth from HST data\tablenotemark{b}		&$>1.6\times10^{-5}$ &$1.0\times10^{-5}$ \\
Total Dust Mass (g)\tablenotemark{c}		&$>7.1\times10^{23}$& $5.2\times10^{23}$ \\

\enddata
%
%
\tablenotetext{a}{The fractional dust luminosity
\citep{zuck04}.
}
\tablenotetext{b}{Derived from the model disks.
Since the disks are optically thin, we sum the cumulative
light from the model disk, $m_d$, and quote optical depth as $10^{(m_d - m_V) / -2.5)}$.  We assume
albedo=1 and the optical depth will scale inversely with the assumed albedo.
}
\tablenotetext{c}{Dust mass follows from the HST optical depth and assumes a uniform particle
radius of 30 $\mu$m, density 2.5 g cm$^{-3}$ and albedo=1.0.  
}

%
\end{deluxetable}
\clearpage

\begin{deluxetable}{lllllll}
\tabletypesize{\scriptsize}
\tablecaption{Debris disk architectures from scattered light properties\tablenotemark{a}
\label{tbl-1}}
\tablewidth{0pt}
\tablehead{
\colhead{Name} & \colhead{$r_{in}$ (AU)} & \colhead{$r_{out}$ (AU)}  & \colhead{Width (AU)} & \colhead{d (pc)} & \colhead{SpT} & \colhead{References}
}
\startdata
$\beta$ Pic	&$\sim$90	&$>$1835  	&  \bf{$>$1745}   & 19.3  &  A5V  &  1, 2\\
HD 32297	&$<$40		&$>$1680  	&  \bf{$>$1640}   & 112   &  A0     &  3, 4\\
AU Mic		&$\sim$12	&$>$210		&  \bf{$>$198}     & 9.9     &  M1Ve &  5, 6\\
HD 53143	&$<$55		&$>$110  	&  \bf{$>$55}   	   & 18.4  &  K2V &  This work.\\
HD 107146	&$\sim$130	&$>$185  	&  \bf{$>$55}        & 28.4  &  G2V &  7\\
&&\\
HD 139664	& 83	& 109	& \bf{26}	& 18.5  &F5V	&  This work.\\
Fomalhaut	& 133  &  158  &  \bf{25}  	&  7.7 	& A3V  &  8\\
HR 4796A	& 60  &  80  &  \bf{20}  	&  67.1 	& A0V  &  9\\
Sun\tablenotemark{b}	& 25  & 50  & \bf{25}  & --  & G2V  & 10, 11
\enddata
%
%
\tablenotetext{a}{We do not include systems younger than 10 Myr that are likely to possess significant primordial circumstellar gas.  
Column 2 gives the inner disk radius corresponding to the approximate 
peak of dust number density.  Column 3 gives the outer radius.
}
\tablenotetext{b}{From simulations of relatively large grains 
trapped in resonances with Neptune.
}

\tablecomments{REFERENCES:  (1) \citet{pantin97}; (2) \citet{larwood01};
(3) \citet{ssh05}; (4) \citet{kalas05b}; (5) \citet{krist05}; (6) \citet{kalas04}; (7) \citet{ardila04};
(8) \citet{kalas05a}; (9) \citet{schneider99}; (10) \citet{liou99};  (11) \citet{moro02}
}
\end{deluxetable}

\end{document}